\documentclass[aps,prl,twocolumn,showpacs,superscriptaddress,groupedaddress]{revtex4}  
\usepackage{graphicx}  
\usepackage{dcolumn}   
\usepackage{bm}        
\usepackage{amssymb}   
\usepackage{cancel}
\usepackage{amsmath}
\usepackage{natbib}
\usepackage{hyperref}

\newcommand{\no}{\rm no\,}
\begin{document}

\title{A First Detection of the Acoustic Oscillation Phase Shift Expected from the Cosmic Neutrino Background}
\author{Brent Follin}
\affiliation{University of California, Davis}
\email{btfollin@ucdavis.edu}
\author{Lloyd Knox}
\affiliation{University of California, Davis}
\author{Marius Millea}
\affiliation{University of California, Davis}
\author{Zhen Pan}
\affiliation{University of California, Davis}

\begin{abstract}

The unimpeded relativistic propagation of cosmological neutrinos prior to recombination of the baryon-photon plasma alters gravitational potentials and therefore the details of the time-dependent gravitational driving of acoustic oscillations.  We report here a first detection of the resulting shifts in the temporal phase of the oscillations, which we infer from their signature in the Cosmic Microwave Background (CMB) temperature power spectrum. 

\end{abstract}

\maketitle

{\it Introduction:} The hot, dense conditions of the early universe included a thermal background of photons. Initially scattering with a high rate off of the free electrons in the plasma, they eventually streamed freely as the plasma cooled to $k_BT \sim 0.3$ eV, the density of free electrons plummeted, and the mean free path became larger than the extent of the observable universe. Today we observe these photons, about 90\% of which last scattered at this epoch, as the cosmic microwave background (CMB), stretched by expansion into millimeter wavelengths.

These same hot and dense conditions led to a cosmic neutrino background (CNB) that contributes nearly as much as photons to the total energy density in the early universe.  The neutrinos began to stream freely  at $k_BT \sim \,$MeV, and continue to stream freely through the cosmos to this day. Unlike with photons, direct detection of the CNB is exceedingly difficult \citep{betts_development_2013}. 

We have had indirect, though highly significant, evidence for the CNB for decades, starting with determinations of the primordial abundance of Helium, an abundance affected by the contribution of neutrinos to the expansion rate (see \citet{steigman_neutrinos_2012}). More recently we have evidence from the amplitude of the damping tail region of the CMB angular power spectrum \citep{hou_how_2013} \citep{planck_collaboration_planck_2015} \citep{WMAP7_cosmological_parameters}. 

Due to the indirect nature of these inferences of the CNB, they are highly model dependent. Therefore we are motivated to find signatures of the CNB that are especially robust to changes in model assumptions. We report here the first detection of a particularly robust, though subtle, effect of the CNB: its influence on the temporal phase of the acoustic oscillations of the primordial plasma. 

\citet{bashinsky_signatures_2004} analytically found that the propagation of neutrino perturbations at speeds faster than the speed of sound in the plasma alters the time-dependent gravitational driving of the acoustic oscillations. Altering the driving of a harmonic oscillator has two effects: it changes the amplitude and the temporal phase of the oscillations. To date, no one has isolated these two distinct effects. Although the influence of neutrino perturbations has been detected \citep{trotta_indication_2005,planck_collaboration_planck_2015}, it remains an open question whether the phase shift plays a significant role in these determinations.

Isolating the phase shift effect clarifies the amount of model dependence in our inferences of the CNB. The amplitude effect has a very smooth dependence on angular scale, and thus can be mimicked by a smooth scale-dependent alteration of the statistical properties of the initial conditions. In contrast, mimicking the phase shift with an alteration to the initial conditions would require an oscillatory scale dependence that just happens to have a period matched to that of the acoustic peaks. It is also very difficult to mimic the phase shift by alterations to the matter content\citep{bashinsky_signatures_2004}, short of substituting some other dark and relativistic species in place of the expected cosmological neutrinos. 

In what follows we use a phenomenological parameterization of the influence of the phase shift on observables to directly determine the size of the phase shift due to the CNB. Although it is a subtle effect, with the {\it Planck} data it is finally possible to unambigously isolate this robust signature of neutrino free streaming.

The power spectrum is $C_\ell \equiv \langle |a_{\ell m}|^2 \rangle$ where $T(\theta,\phi) = \sum_{\ell m} a_{\ell m} Y_{\ell m}(\theta,\phi)$ is a decomposition of the temperature map into spherical harmonics and $\langle ... \rangle$ indicates an ensemble average. A $Y_{\ell m}$ undergoes $\ell$ oscillations every 360 degrees, so, e.g., $Y_{180 m}$ are patterns with hot and cold spots separated by about $1^\circ$. We show model and measured power spectra in Fig. 1. 

\begin{figure*}
\includegraphics[scale=0.8]{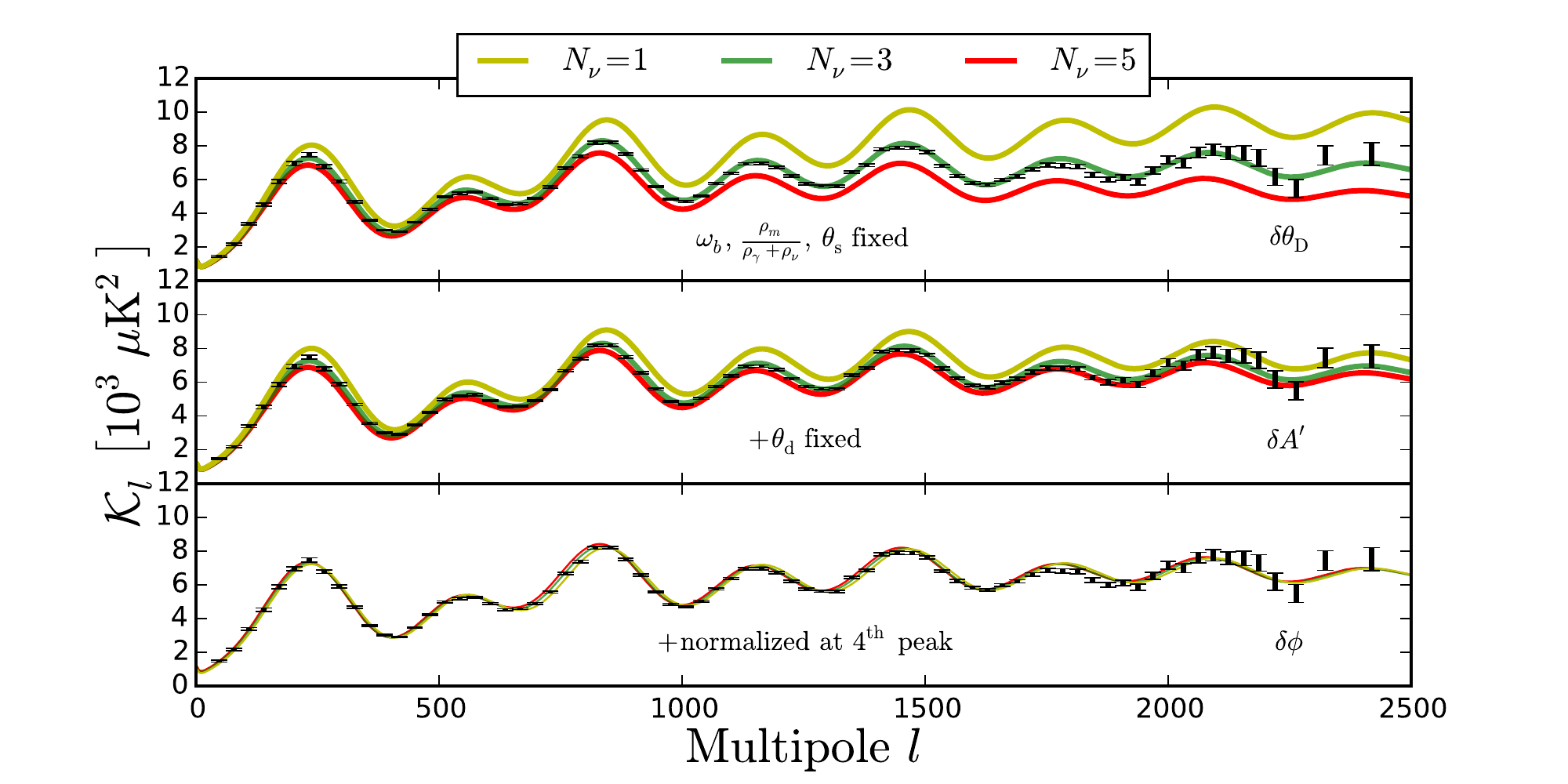}
\caption{\label{fig:powerspectra}  Undamped power spectra ${\cal K}_l$ (defined in Eq.~\ref{eq:kell}) with different values of $N_\nu$. In all panels the baryon density $\omega_b$, the ratio of matter to radiation density $\rho_m/(\rho_\gamma + \rho_\nu)$ and the
angular size of the sound horizon $\theta_s$ are held fixed as these are well-determined by CMB data fairly independently of the assumed value of $N_\nu$. In the top panel the dominant source of variation
is the change in the damping scale $\theta_D$ caused by the changes in $N_\nu$. In the middle panel we fix $\theta_D$ by varying the primordial fraction of baryonic mass in Helium
appropriately, leaving the dominant source of power spectrum variation as the change in oscillation amplitude $A^\prime$. Finally in the bottom panel, with the spectra normalized to remove the
effect of $A^\prime$ variation, one can see the subtle impact of the shifts in temporal phase $\phi$. The data points are the 2013 Planck data.}
\end{figure*}

The series of peaks is what we expect from a collection of standing waves, all beginning their amplitude oscillations with zero initial momentum\footnote{These are the ``initial conditions'' generated by the simplest models of inflation.}. For example, the Fourier mode with spatial dependence $\delta \rho_\gamma \propto \cos(\vec k \cdot \vec x)$ will have a time dependent amplitude approximately given by $A \cos[k r_s(t)]$, where $r_s(t)$ is the distance a sound wave can travel by time $t$ (and we have ignored, for now, a time-dependent driving effect that alters this simple-harmonic motion solution). Modes with (comoving) wavelengths of $\sim 400$ Mpc project into $\ell \sim k d = \frac{2\pi}{400 {\rm \ Mpc}}d = 220$ where $d$ is the (comoving angular-diameter) distance to the last-scattering surface. These modes oscillate with a period such that they achieve their first extremum at the epoch of last scattering, $t=t_*$, and thus produce the first peak in the power spectrum.  Modes with a wavelength of $\sim 220$ Mpc project into $\ell \simeq 400$. They have a shorter oscillation period so that by $t=t_*$ they have gone through their first extremum and reached a null at the time of decoupling. Modes that hit their $p^{\rm th}$ extremum at decoupling contribute to the $p^{\rm th}$ peak. 

A key length scale for understanding the response of $C_\ell$ to the CNB is the sound horizon at decoupling, $r_s(t_*)$. The sound horizon is smaller than it would be without the presence of the CNB, because the fractional expansion rate $H \propto \bar \rho_{\rm tot}^{1/2}$. If we scaled up the fractional expansion rate $H$ at all times by a factor $\alpha$, the decrease in time it takes for the temperature to drop to $T \simeq 0.3$ eV would lead to $r_s \propto 1/\alpha$.  The angle $r_s$ subtends, $\theta_s = r_s(t_*)/d$ where $d$ is the angular-diameter distance to the last-scattering surface, strongly influences the locations of the acoustic peaks such that $\delta \ell_p = \ell_p \delta \theta_s/\theta_s$. If we knew $d$, we could use this effect alone to measure the energy density in the CNB.  However, $d$ depends on the (otherwise unknown) value of the cosmological constant.

In Fig.~\ref{fig:powerspectra} we show a series of plots where we vary $N_{\nu}$ while holding certain other quantities fixed, in order to demonstrate the observable consequences of various effects of neutrinos.  Because $\theta_s$, baryon density $\omega_b$, and the ratio of matter to radiation density $\rho_m/(\rho_\gamma + \rho_\nu)$ are well determined by the data, in all rows we show variations with these parameters fixed. In the top row one can see the impact of $N_{\nu}$ on the typical distance a photon diffuses prior to last scattering, $r_d$.
This diffusion suppresses anisotropy for modes with wavelengths $\lambda \lesssim r_D$, with an approximate effect of $C_\ell \rightarrow D_\ell C_\ell$ where 
$D_\ell \simeq \exp\left[-(\ell \theta_D)^{1.18}\right]$ where $\theta_D = r_D/d$. Because the diffusion is a stochastic process it scales
with expansion rate as $1/\sqrt{\alpha}$ rather than $1/\alpha$ as $r_s$ does. These different scalings mean that while we adjust $d$
to keep $\theta_s$ fixed, we get $\theta_D \propto \sqrt{\alpha}$. Thus for $N_\nu$ = 5, the expansion rate is greater, leading to larger $\theta_D$ and more damping.

To visualize more subtle effects of the CNB, we can vary the priordial fraction of baryonic mass in Helium, $Y_p$, to keep $\theta_D$ fixed as well \citep{bashinsky_signatures_2004}. Doing so in the middle panel, we can see an impact of the perturbations in the CNB.  As an initially over-dense region compresses under the influence of gravity, the compression does not occur rapidly enough to prevent the gravitational potential from decaying due to the expansion-driven drop in density. By the time pressure gradients halt the compression, the gravitational potential has nearly completely decayed. The result of this temporary time-dependent gravitational driving of the accoustic oscillations is a change to the subsequent amplitude and phase so that the amplitude of a standing wave is $A' \cos[k r_s(t)+ \phi]$. The values of $A'/A$ and $\phi$ depend on 
the details of the potential decay, and in particular on the fraction of the radiation that can freely stream out of over densities at the speed of light. We can see in the middle row that increasing $N_\nu$ decreases $A'/A$.  

Finally, in the bottom row we normalize the spectra to remove the effect of changing $A'/A$ so that the change in values of $\phi$ is more evident. As $\phi$ changes, the value of $k=k_p$ for which $k_pr_s(t_*)+\phi = p\pi$ changes so $\ell_p \simeq k_p d$ changes. The net result is $\delta \ell_p = -\delta \phi/\theta_s$. In this {\em Letter} we show that these subtle shifts are detectable with the {\it Planck} data.

{\it Template fitting:} 
To quantify the sensitivity of the data to the expected phase shift we must be able to artificially increase or decrease it, independent of other effects of the CNB. Since the phase shift effects are most observable deep in the damping tail, we work with an (approximately) undampted spectrum 
\begin{equation}
\label{eq:kell}
{\cal K}_\ell \equiv \ell(\ell + 1)/(2\pi)C_\ell D_\ell^{-1},
\end{equation}
where $ D_\ell^{-1}$ approximately undoes the damping expected for three neutrino species and standard Big Bang Nucleosynthesis. We then define a transformation that sends ${\cal K}_\ell  \rightarrow {\cal K}_{\ell + \delta \ell_\nu}$, which is controlled by a new parameter $N_\nu^{\delta \phi}$. Following the analytic work of \citet{bashinsky_signatures_2004} we take the amplitude of $\delta \ell_\nu$ to
be linear in the fraction of radiation energy density in neutrinos, $R(N_\nu)$, so that
\begin{align}
\delta \ell_\nu & = A(N_{\nu}^{\delta \phi}, N_{\nu}) f_\ell,
\end{align}
with $N_{\nu}^{\delta \phi}$ defined such that  
\begin{equation}
A(N_{\nu}^{\delta \phi}, N_{\nu}) \propto R(N_{\nu}^{\delta \phi}) - R(N_{\nu}),
\label{eq:phiamp}
\end{equation}
with proportionality chosen to match the shift from the standard $3.046$ to one neutrino specie. 

The analytic result of \citep{bashinsky_signatures_2004} was obtained by ignoring the presence of any pressureless matter, an approximation that becomes increasingly valid as one considers modes that began oscillating deeper in the radiation-dominated era. With this approximation they find a $\delta \phi$ (and therefore $f_\ell$ in our language) that is independent of $k$ and therefore independent of $\ell$. To avoid this approximation we have numerically calculated $\delta\ell_\nu$ from the $C_\ell$ calculated with the Boltzmann equation solver CLASS \citep{blas_cosmic_2011} for a suite of models with varying $N_\nu$. To define the suite of models
we compare two cosmologies: one from the $\Lambda$CDM posterior of the $2013$ {\it Planck} data, and the other with nonstandard $N_\nu$ drawn from $1 < N_\nu < 6$. We isolate the effects of neutrinos from the effects of the other components contributing to gravitational potentials, as well as effects of changing $N_\nu$ to the background Hubble rate $H(z)$, by following \citet{hou_how_2013} in fixing the baryon density $\omega_b$, redshift of matter-radiation equality $z_{\rm eq}$, Silk damping scale $\theta_D$, and sound horizon scale $\theta_s$, as well as the spectral amplitude and tilt describing the adiabatic initial conditions from inflation between the two models. 
Since contributions to anisotropy arising after recombination project differently, we zero out contributions from the integrated Sachs Wolfe (ISW) effect.  

\begin{figure}
\includegraphics[scale=0.6]{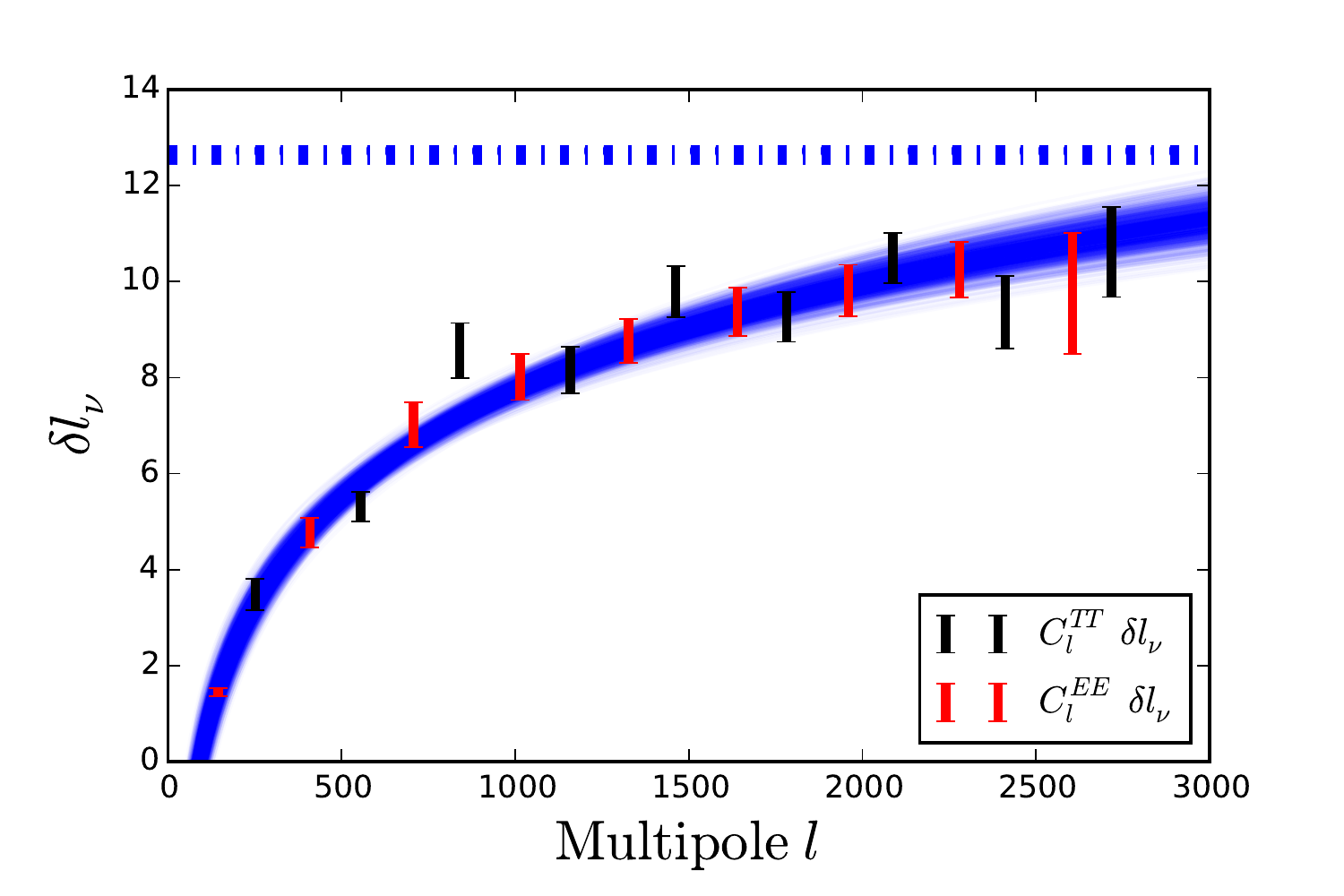}
\caption{\label{fig:template} Posterior reconstructions of the phase shift template $f_l=\alpha \ln(l) + \beta$ under the $\Lambda$CDM + $N_{\nu}$ + $N_{\nu}^{\delta \phi}$ model, normalized to give the amplitude of the relative shift between the fiducial 3.046 neutrino species and a single specie, along with numerically obtained phase shifts obtained from 100 different simulated cosmologies with varying neutrino species drawn from $N_{\nu} = \left[1,6\right]$ relative to the fiducial cosmology, rescaled by the amplitude parameter given in equation \ref{eq:phiamp}.  Also included is the analytic approximation given by \citep{bashinsky_signatures_2004}, which they also found to be $\sim 25\%$ higher than the numerical result at $\ell \simeq 3000$.}  
\end{figure}

We sample 100 different cosmology pairs, and find that in the region of parameter space explored by these models, the phase shift is well captured by a linear response proportional to the fraction of radiation density in free-streaming neutrinos.
The multipole dependence is well described by a logarithmic template, which is jointly sampled with cosmology against both the March 2013 {\it Planck} temperature data and the measured phase shifts in the 100 cosmology pairs.
Posterior samples of the template are shown in Fig.~\ref{fig:template}, along with the measured phase shifts $\delta l_\nu$ at the peaks in the ISW-less temperature power spectrum ${\cal K}_l^{TT}$, as well as, for visualization, the locations of the corresponding peaks in the polarization spectrum ${\cal K}_l^{EE}$, which is not used in the fit.

To apply this template to models we confront with the data, we first decompose the temperature power spectrum $C_l$ into ISW and ISW-less components: $C_l = C_l^{\no {\rm ISW}} + C_l^{\rm ISW} + C_l^{\rm cross}$,
with $C_l^{\rm cross}$ the (ISW) $\times$ (no ISW) contribution.
The artificial $\ell-$space shift we introduce is given by 
\begin{align}
\label{eq:phaseshift}
{\cal K}_\ell &\rightarrow {\cal K}_{\ell + \delta \ell_\nu}^{\no {\rm ISW}} + {\cal K}_l^{\rm ISW} + {\cal K}_l^{\rm cross} \\ \nonumber
\delta \ell_\nu & = A(N_{\nu}^{\delta \phi}, N_{\nu}) f(\ell),
\end{align}
with the implied definitions for the various ${\cal K}_\ell$ components. The artificial $\ell-$space shift alters the power spectrum away from that of the physical model which always has $N_\nu = N_\nu^{\delta \phi}$, for which $A=0$. To vary just the phase shift effect, we can set $N_{\nu} = 3.046$, the fiducial value, while varying $N_{\nu}^{\delta \phi}$. 

{\it Results from $Planck$}: We use the publically available likelihood code {\bf clik} \citep{planck_collaboration_planck_2014-1} to determine constraints from the 2013 $Planck$ temperature power spectrum measurements, with the polarization constraints approximated as a Gaussian prior on the optical depth to last scattering $\tau$ for simplicity\footnote{A Fisher matrix analysis shows that using the $\tau$ prior instead of the ``WP'' of \citet{planck_collaboration_planck_2014-1} does not lose any information relevant to the phase shift}. 
We place uniform (flat) priors on the parameters $N_{\nu}$ and $N_{\nu}^{\delta\phi}$, which results in a flat prior for the physical case where $N_{\nu} = N_{\nu}^{\delta \phi}$.  We explore the model space using the MCMC routines provided by the Python library {\texttt CosmoSlik} \footnote{ {\url https://github.com/marius311/cosmoslik.git}}. 

\begin{figure}
\includegraphics[scale=0.6]{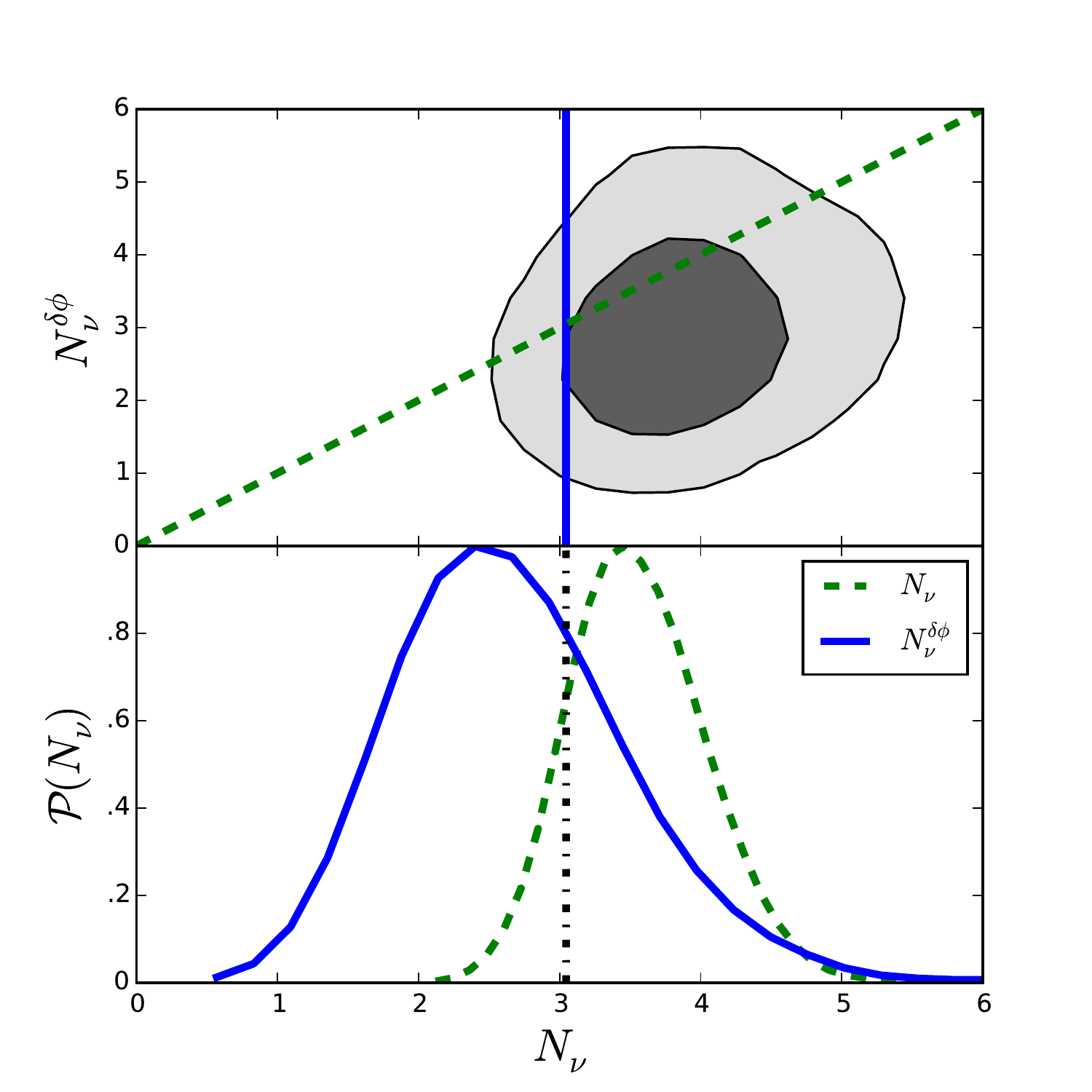}
\caption{\label{fig:nphase1d} 
{\bf Top}: 2D constraints on the jointly varying $\Lambda$CDM+$N_{\nu}$+$N_{\nu}^{\delta\phi}$ parameter space. The constraints on $N_{\nu}$ (damping) and $N_{\nu}^{\delta\phi}$ (phase shift) are essentially orthogonal. 
{\bf Bottom}: Constraints from March 2013 ${\it Planck}$ temperature power spectrum measurements on the number of neutrino species from (1) {\it blue/solid:} varying $N_{\nu}^{\delta\phi}$ while holding $N_{\nu}$ fixed at three and (2) {\it green/dashed:} varying along the physical direction $N_{\nu} = N_{\nu}^{\delta\phi}$. The constraints assume a Gaussian $\tau$ prior of mean $\mu = 0.085$ and width $\sigma = 0.015$.}
\end{figure}

In Fig.~\ref{fig:nphase1d} we show the constraint on $N_{\nu}^{\delta\phi}$ from the $\Lambda$CDM $+N_\nu^{\delta\phi}$ model.  For comparison, we include the constraints on the $\Lambda$CDM +$N_{\nu}$ model space, with the phenomenological amplitude $A(N_\nu^{\delta\phi}, N_{\nu}) \equiv 0$. We find best-fit values of $N_\nu = 3.3^{+0.7}_{-0.2}$ \footnote{This approximation results in a $\sim 10\%$ widening of the constraint on $N_{\nu}$ in the standard model extended to arbitrary number of neutrino species when compared to the full CMB likelihood, which includes WMAP$9$ polarization.} and $N_{\nu}^{\delta \phi} = 2.3^{+1.1}_{-0.4}$. To quantify the significance
of the detection we conducted a likelihood ratio test. We found that the minimum $\chi^2$ ($\equiv - 2 \ln L$) decreased by 19.9 when switching from a $\Lambda$CDM model with $N_\nu^{\delta \phi} = 0$ to the $\Lambda$CDM + $N_\nu^{\delta \phi}$ model. A $\chi^2$ difference this large or larger will occur, assuming $N_\nu^{\delta \phi} = 0$, with the same probability,  $8 \times 10^{-6}$, as a 4.5$\sigma$ Gaussian fluctuation.

While letting $N_\nu$ and $N_\nu^{\delta\phi}$ vary independently (top of Fig.~\ref{fig:nphase1d}), the width of the constraint on $N_\nu$ is dominated by an $n_s$- $N_\nu$ degeneracy: fixing $n_s$ results in roughly a halving of the characteristic width of the posterior in the $N_{\nu}$ direction.  For the $N_\nu^{\delta \phi}$ direction, no such correlations exist, so while our constraints on $N_\nu$ depend somewhat on the characterization of initial conditions due to inflation (as well as, in extended models such one with the helium fraction $Y_p$ free), our constraint on $N_\nu^{\delta \phi}$ is due to a feature in the data that is difficult to mimic with other cosmological degrees of freedom. In addition, we find the constraints on $N_\nu$ and $N_\nu^{\delta \phi}$ are nearly uncorrelated. This lack of correlation follows from the fact that the response of $C_l$ to changing the phase shift is essentially orthogonal to the response due to other observable effects of the cosmic neutrino background in the CMB. 

Finally, we note that there is a slight dependence on priors for the 2D posterior shown in Fig.~\ref{fig:nphase1d}.  If we switch from uniform priors on the number of neutrino species $N_{\nu}$ and $N_{\nu}^{\delta\phi}$ to their corresponding neutrino fractions $R(N_{\nu})$ and $R(N_{\nu}^{\delta\phi})$, the average value of the posterior shifts down by $\left(\Delta N_{\nu} = 0.3,\Delta N_\nu^{\delta\phi} = 0.5\right)$, a shift of slightly more than $0.5\sigma$ in both directions.  This is predominantly due to a contraction in the high probability region at high $N_{\nu}$ or $N_{\nu}^{\delta \phi}$.
Regardless of prior, $N_{\nu}^{\delta \phi} = 0$ is heavily disfavored.

{\it Conclusions:}
In this letter, we present the first detection of the temporal phase shift generated by neutrino perturbations during the acoustic oscillation phase of cosmological evolution, and find an amplitude of this effect consistent with the standard value associated with the three known neutrino species.  As pointed out by \citet{bashinsky_signatures_2004} this subtle signature is a particularly robust one, difficult to mimic by either changes to the initial conditions or matter content. Our detection of this effect is thus the most model-independent determination to date of the existence of the CNB. CMB Polarization measurements will bring even more robustness as they are capable
of definitively ruling out the remote possibility of mimicking the phase shift effect by a designer alteration of the initial conditions. 

Were we to have found $N_\nu \ne N_\nu^{\delta \phi}$ we would be compelled to look for a physical explanation, such as $\nu$ $\nu$ scattering which could inhibit free streaming \citep{CyrRacine_2014_limits}\footnote{For a summary of existing constraints on ``secret neutrino interactions'' see \citet{ng_2015_cascades}.}. With current data, we see consistency with the standard model. Future datasets, such as from $Planck$ including polarization spectra \citep{the_planck_collaboration_scientific_2006} and CMB stage-4 power spectra \citep{abazajian_neutrino_2013} will provide even stronger constraints on the phase shift, either providing a signature of new physics or increasing confidence in the standard cosmological model. We forecast that these will achieve $\sigma(N_\nu^{\delta \phi}) = 0.41$ and 0.09 respectively.

\bibliographystyle{apsrev4-1} 
\bibliography{Neffphase}

\begin{thebibliography}{13}%
\makeatletter
\providecommand \@ifxundefined [1]{%
 \@ifx{#1\undefined}
}%
\providecommand \@ifnum [1]{%
 \ifnum #1\expandafter \@firstoftwo
 \else \expandafter \@secondoftwo
 \fi
}%
\providecommand \@ifx [1]{%
 \ifx #1\expandafter \@firstoftwo
 \else \expandafter \@secondoftwo
 \fi
}%
\providecommand \natexlab [1]{#1}%
\providecommand \enquote  [1]{``#1''}%
\providecommand \bibnamefont  [1]{#1}%
\providecommand \bibfnamefont [1]{#1}%
\providecommand \citenamefont [1]{#1}%
\providecommand \href@noop [0]{\@secondoftwo}%
\providecommand \href [0]{\begingroup \@sanitize@url \@href}%
\providecommand \@href[1]{\@@startlink{#1}\@@href}%
\providecommand \@@href[1]{\endgroup#1\@@endlink}%
\providecommand \@sanitize@url [0]{\catcode `\\12\catcode `\$12\catcode
  `\&12\catcode `\#12\catcode `\^12\catcode `\_12\catcode `\%12\relax}%
\providecommand \@@startlink[1]{}%
\providecommand \@@endlink[0]{}%
\providecommand \url  [0]{\begingroup\@sanitize@url \@url }%
\providecommand \@url [1]{\endgroup\@href {#1}{\urlprefix }}%
\providecommand \urlprefix  [0]{URL }%
\providecommand \Eprint [0]{\href }%
\providecommand \doibase [0]{http://dx.doi.org/}%
\providecommand \selectlanguage [0]{\@gobble}%
\providecommand \bibinfo  [0]{\@secondoftwo}%
\providecommand \bibfield  [0]{\@secondoftwo}%
\providecommand \translation [1]{[#1]}%
\providecommand \BibitemOpen [0]{}%
\providecommand \bibitemStop [0]{}%
\providecommand \bibitemNoStop [0]{.\EOS\space}%
\providecommand \EOS [0]{\spacefactor3000\relax}%
\providecommand \BibitemShut  [1]{\csname bibitem#1\endcsname}%
\let\auto@bib@innerbib\@empty
\bibitem [{\citenamefont {Betts}\ \emph {et~al.}(2013)\citenamefont {Betts},
  \citenamefont {Blanchard}, \citenamefont {Carnevale}, \citenamefont {Chang},
  \citenamefont {Chen}, \citenamefont {Chidzik}, \citenamefont {Ciebiera},
  \citenamefont {Cloessner}, \citenamefont {Cocco}, \citenamefont {Cohen},
  \citenamefont {Dong}, \citenamefont {Klemmer}, \citenamefont {Komor},
  \citenamefont {Gentile}, \citenamefont {Harrop}, \citenamefont {Hopkins},
  \citenamefont {Jarosik}, \citenamefont {Mangano}, \citenamefont {Messina},
  \citenamefont {Osherson}, \citenamefont {Raitses}, \citenamefont {Sands},
  \citenamefont {Schaefer}, \citenamefont {Taylor}, \citenamefont {Tully},
  \citenamefont {Woolley},\ and\ \citenamefont
  {Zwicker}}]{betts_development_2013}%
  \BibitemOpen
  \bibfield  {author} {\bibinfo {author} {\bibfnamefont {S.}~\bibnamefont
  {Betts}}, \bibinfo {author} {\bibfnamefont {W.~R.}\ \bibnamefont
  {Blanchard}}, \bibinfo {author} {\bibfnamefont {R.~H.}\ \bibnamefont
  {Carnevale}}, \bibinfo {author} {\bibfnamefont {C.}~\bibnamefont {Chang}},
  \bibinfo {author} {\bibfnamefont {C.}~\bibnamefont {Chen}}, \bibinfo {author}
  {\bibfnamefont {S.}~\bibnamefont {Chidzik}}, \bibinfo {author} {\bibfnamefont
  {L.}~\bibnamefont {Ciebiera}}, \bibinfo {author} {\bibfnamefont
  {P.}~\bibnamefont {Cloessner}}, \bibinfo {author} {\bibfnamefont
  {A.}~\bibnamefont {Cocco}}, \bibinfo {author} {\bibfnamefont
  {A.}~\bibnamefont {Cohen}}, \bibinfo {author} {\bibfnamefont
  {J.}~\bibnamefont {Dong}}, \bibinfo {author} {\bibfnamefont {R.}~\bibnamefont
  {Klemmer}}, \bibinfo {author} {\bibfnamefont {M.}~\bibnamefont {Komor}},
  \bibinfo {author} {\bibfnamefont {C.}~\bibnamefont {Gentile}}, \bibinfo
  {author} {\bibfnamefont {B.}~\bibnamefont {Harrop}}, \bibinfo {author}
  {\bibfnamefont {A.}~\bibnamefont {Hopkins}}, \bibinfo {author} {\bibfnamefont
  {N.}~\bibnamefont {Jarosik}}, \bibinfo {author} {\bibfnamefont
  {G.}~\bibnamefont {Mangano}}, \bibinfo {author} {\bibfnamefont
  {M.}~\bibnamefont {Messina}}, \bibinfo {author} {\bibfnamefont
  {B.}~\bibnamefont {Osherson}}, \bibinfo {author} {\bibfnamefont
  {Y.}~\bibnamefont {Raitses}}, \bibinfo {author} {\bibfnamefont
  {W.}~\bibnamefont {Sands}}, \bibinfo {author} {\bibfnamefont
  {M.}~\bibnamefont {Schaefer}}, \bibinfo {author} {\bibfnamefont
  {J.}~\bibnamefont {Taylor}}, \bibinfo {author} {\bibfnamefont {C.~G.}\
  \bibnamefont {Tully}}, \bibinfo {author} {\bibfnamefont {R.}~\bibnamefont
  {Woolley}}, \ and\ \bibinfo {author} {\bibfnamefont {A.}~\bibnamefont
  {Zwicker}},\ }\href {http://arxiv.org/abs/1307.4738} {\bibfield  {journal}
  {\bibinfo  {journal} {arXiv:1307.4738 [astro-ph, physics:physics]}\ }
  (\bibinfo {year} {2013})},\ \bibinfo {note} {arXiv: 1307.4738}\BibitemShut
  {NoStop}%
\bibitem [{\citenamefont {Steigman}(2012)}]{steigman_neutrinos_2012}%
  \BibitemOpen
  \bibfield  {author} {\bibinfo {author} {\bibfnamefont {G.}~\bibnamefont
  {Steigman}},\ }\href {\doibase 10.1155/2012/268321} {\bibfield  {journal}
  {\bibinfo  {journal} {Advances in High Energy Physics}\ }\textbf {\bibinfo
  {volume} {2012}},\ \bibinfo {pages} {1} (\bibinfo {year} {2012})},\ \bibinfo
  {note} {arXiv: 1208.0032}\BibitemShut {NoStop}%
\bibitem [{\citenamefont {Hou}\ \emph {et~al.}(2013)\citenamefont {Hou},
  \citenamefont {Keisler}, \citenamefont {Knox}, \citenamefont {Millea},\ and\
  \citenamefont {Reichardt}}]{hou_how_2013}%
  \BibitemOpen
  \bibfield  {author} {\bibinfo {author} {\bibfnamefont {Z.}~\bibnamefont
  {Hou}}, \bibinfo {author} {\bibfnamefont {R.}~\bibnamefont {Keisler}},
  \bibinfo {author} {\bibfnamefont {L.}~\bibnamefont {Knox}}, \bibinfo {author}
  {\bibfnamefont {M.}~\bibnamefont {Millea}}, \ and\ \bibinfo {author}
  {\bibfnamefont {C.}~\bibnamefont {Reichardt}},\ }\href {\doibase
  10.1103/PhysRevD.87.083008} {\bibfield  {journal} {\bibinfo  {journal}
  {Physical Review D}\ }\textbf {\bibinfo {volume} {87}},\ \bibinfo {pages}
  {083008} (\bibinfo {year} {2013})}\BibitemShut {NoStop}%
\bibitem [{\citenamefont {{The Planck
  Collaboration}}(2015)}]{planck_collaboration_planck_2015}%
  \BibitemOpen
  \bibfield  {author} {\bibinfo {author} {\bibnamefont {{The Planck
  Collaboration}}},\ }\href {http://arxiv.org/abs/1502.01589} {\bibfield
  {journal} {\bibinfo  {journal} {arXiv:1502.01589 [astro-ph]}\ } (\bibinfo
  {year} {2015})},\ \bibinfo {note} {arXiv: 1502.01589}\BibitemShut {NoStop}%
\bibitem [{\citenamefont {{Komatsu}}\ \emph {et~al.}(2011)\citenamefont
  {{Komatsu}}, \citenamefont {{Smith}}, \citenamefont {{Dunkley}},
  \citenamefont {{Bennett}}, \citenamefont {{Gold}}, \citenamefont {{Hinshaw}},
  \citenamefont {{Jarosik}}, \citenamefont {{Larson}}, \citenamefont {{Nolta}},
  \citenamefont {{Page}}, \citenamefont {{Spergel}}, \citenamefont {{Halpern}},
  \citenamefont {{Hill}}, \citenamefont {{Kogut}}, \citenamefont {{Limon}},
  \citenamefont {{Meyer}}, \citenamefont {{Odegard}}, \citenamefont {{Tucker}},
  \citenamefont {{Weiland}}, \citenamefont {{Wollack}},\ and\ \citenamefont
  {{Wright}}}]{WMAP7_cosmological_parameters}%
  \BibitemOpen
  \bibfield  {author} {\bibinfo {author} {\bibfnamefont {E.}~\bibnamefont
  {{Komatsu}}}, \bibinfo {author} {\bibfnamefont {K.~M.}\ \bibnamefont
  {{Smith}}}, \bibinfo {author} {\bibfnamefont {J.}~\bibnamefont {{Dunkley}}},
  \bibinfo {author} {\bibfnamefont {C.~L.}\ \bibnamefont {{Bennett}}}, \bibinfo
  {author} {\bibfnamefont {B.}~\bibnamefont {{Gold}}}, \bibinfo {author}
  {\bibfnamefont {G.}~\bibnamefont {{Hinshaw}}}, \bibinfo {author}
  {\bibfnamefont {N.}~\bibnamefont {{Jarosik}}}, \bibinfo {author}
  {\bibfnamefont {D.}~\bibnamefont {{Larson}}}, \bibinfo {author}
  {\bibfnamefont {M.~R.}\ \bibnamefont {{Nolta}}}, \bibinfo {author}
  {\bibfnamefont {L.}~\bibnamefont {{Page}}}, \bibinfo {author} {\bibfnamefont
  {D.~N.}\ \bibnamefont {{Spergel}}}, \bibinfo {author} {\bibfnamefont
  {M.}~\bibnamefont {{Halpern}}}, \bibinfo {author} {\bibfnamefont {R.~S.}\
  \bibnamefont {{Hill}}}, \bibinfo {author} {\bibfnamefont {A.}~\bibnamefont
  {{Kogut}}}, \bibinfo {author} {\bibfnamefont {M.}~\bibnamefont {{Limon}}},
  \bibinfo {author} {\bibfnamefont {S.~S.}\ \bibnamefont {{Meyer}}}, \bibinfo
  {author} {\bibfnamefont {N.}~\bibnamefont {{Odegard}}}, \bibinfo {author}
  {\bibfnamefont {G.~S.}\ \bibnamefont {{Tucker}}}, \bibinfo {author}
  {\bibfnamefont {J.~L.}\ \bibnamefont {{Weiland}}}, \bibinfo {author}
  {\bibfnamefont {E.}~\bibnamefont {{Wollack}}}, \ and\ \bibinfo {author}
  {\bibfnamefont {E.~L.}\ \bibnamefont {{Wright}}},\ }\href {\doibase
  10.1088/0067-0049/192/2/18} {\bibfield  {journal} {\bibinfo  {journal}
  {\apj}\ }\textbf {\bibinfo {volume} {192}},\ \bibinfo {eid} {18} (\bibinfo
  {year} {2011})},\ \Eprint {http://arxiv.org/abs/1001.4538} {arXiv:1001.4538
  [astro-ph.CO]} \BibitemShut {NoStop}%
\bibitem [{\citenamefont {Bashinsky}\ and\ \citenamefont
  {Seljak}(2004)}]{bashinsky_signatures_2004}%
  \BibitemOpen
  \bibfield  {author} {\bibinfo {author} {\bibfnamefont {S.}~\bibnamefont
  {Bashinsky}}\ and\ \bibinfo {author} {\bibfnamefont {U.}~\bibnamefont
  {Seljak}},\ }\href {\doibase 10.1103/PhysRevD.69.083002} {\bibfield
  {journal} {\bibinfo  {journal} {Physical Review D}\ }\textbf {\bibinfo
  {volume} {69}} (\bibinfo {year} {2004}),\ 10.1103/PhysRevD.69.083002},\
  \bibinfo {note} {arXiv: astro-ph/0310198}\BibitemShut {NoStop}%
\bibitem [{\citenamefont {Trotta}\ and\ \citenamefont
  {Melchiorri}(2005)}]{trotta_indication_2005}%
  \BibitemOpen
  \bibfield  {author} {\bibinfo {author} {\bibfnamefont {R.}~\bibnamefont
  {Trotta}}\ and\ \bibinfo {author} {\bibfnamefont {A.}~\bibnamefont
  {Melchiorri}},\ }\href {\doibase 10.1103/PhysRevLett.95.011305} {\bibfield
  {journal} {\bibinfo  {journal} {Physical Review Letters}\ }\textbf {\bibinfo
  {volume} {95}} (\bibinfo {year} {2005}),\ 10.1103/PhysRevLett.95.011305},\
  \bibinfo {note} {arXiv: astro-ph/0412066}\BibitemShut {NoStop}%
\bibitem [{\citenamefont {Blas}\ \emph {et~al.}(2011)\citenamefont {Blas},
  \citenamefont {Lesgourgues},\ and\ \citenamefont {Tram}}]{blas_cosmic_2011}%
  \BibitemOpen
  \bibfield  {author} {\bibinfo {author} {\bibfnamefont {D.}~\bibnamefont
  {Blas}}, \bibinfo {author} {\bibfnamefont {J.}~\bibnamefont {Lesgourgues}}, \
  and\ \bibinfo {author} {\bibfnamefont {T.}~\bibnamefont {Tram}},\ }\href
  {http://arxiv.org/abs/1104.2933} {\bibfield  {journal} {\bibinfo  {journal}
  {arXiv:1104.2933 [astro-ph]}\ } (\bibinfo {year} {2011})},\ \bibinfo {note}
  {arXiv: 1104.2933}\BibitemShut {NoStop}%
\bibitem [{\citenamefont {{The Planck
  Collaboration}}(2014)}]{planck_collaboration_planck_2014-1}%
  \BibitemOpen
  \bibfield  {author} {\bibinfo {author} {\bibnamefont {{The Planck
  Collaboration}}},\ }\href {\doibase 10.1051/0004-6361/201321529} {\bibfield
  {journal} {\bibinfo  {journal} {Astronomy \& Astrophysics}\ }\textbf
  {\bibinfo {volume} {571}},\ \bibinfo {pages} {A1} (\bibinfo {year} {2014})},\
  \bibinfo {note} {arXiv: 1303.5062}\BibitemShut {NoStop}%
\bibitem [{\citenamefont {{Cyr-Racine}}\ and\ \citenamefont
  {{Sigurdson}}(2014)}]{CyrRacine_2014_limits}%
  \BibitemOpen
  \bibfield  {author} {\bibinfo {author} {\bibfnamefont {F.-Y.}\ \bibnamefont
  {{Cyr-Racine}}}\ and\ \bibinfo {author} {\bibfnamefont {K.}~\bibnamefont
  {{Sigurdson}}},\ }\href {\doibase 10.1103/PhysRevD.90.123533} {\bibfield
  {journal} {\bibinfo  {journal} {\prd}\ }\textbf {\bibinfo {volume} {90}},\
  \bibinfo {eid} {123533} (\bibinfo {year} {2014})},\ \Eprint
  {http://arxiv.org/abs/1306.1536} {arXiv:1306.1536} \BibitemShut {NoStop}%
\bibitem [{\citenamefont {{The Planck
  Collaboration}}(2006)}]{the_planck_collaboration_scientific_2006}%
  \BibitemOpen
  \bibfield  {author} {\bibinfo {author} {\bibnamefont {{The Planck
  Collaboration}}},\ }\href {http://arxiv.org/abs/astro-ph/0604069} {\bibfield
  {journal} {\bibinfo  {journal} {arXiv:astro-ph/0604069}\ } (\bibinfo {year}
  {2006})},\ \bibinfo {note} {arXiv: astro-ph/0604069}\BibitemShut {NoStop}%
\bibitem [{\citenamefont {Abazajian}\ \emph {et~al.}(2013)\citenamefont
  {Abazajian}, \citenamefont {Arnold}, \citenamefont {Austermann},
  \citenamefont {Benson}, \citenamefont {Bischoff}, \citenamefont {Bock},
  \citenamefont {Bond}, \citenamefont {Borrill}, \citenamefont {Calabrese},
  \citenamefont {Carlstrom}, \citenamefont {Carvalho}, \citenamefont {Chang},
  \citenamefont {Chiang}, \citenamefont {Church}, \citenamefont {Cooray},
  \citenamefont {Crawford}, \citenamefont {Dawson}, \citenamefont {Das},
  \citenamefont {Devlin}, \citenamefont {Dobbs}, \citenamefont {Dodelson},
  \citenamefont {Dore}, \citenamefont {Dunkley}, \citenamefont {Errard},
  \citenamefont {Fraisse}, \citenamefont {Gallicchio}, \citenamefont
  {Halverson}, \citenamefont {Hanany}, \citenamefont {Hildebrandt},
  \citenamefont {Hincks}, \citenamefont {Hlozek}, \citenamefont {Holder},
  \citenamefont {Holzapfel}, \citenamefont {Honscheid}, \citenamefont {Hu},
  \citenamefont {Hubmayr}, \citenamefont {Irwin}, \citenamefont {Jones},
  \citenamefont {Kamionkowski}, \citenamefont {Keating}, \citenamefont
  {Keisler}, \citenamefont {Knox}, \citenamefont {Komatsu}, \citenamefont
  {Kovac}, \citenamefont {Kuo}, \citenamefont {Lawrence}, \citenamefont {Lee},
  \citenamefont {Leitch}, \citenamefont {Linder}, \citenamefont {Lubin},
  \citenamefont {McMahon}, \citenamefont {Miller}, \citenamefont {Newburgh},
  \citenamefont {Niemack}, \citenamefont {Nguyen}, \citenamefont {Nguyen},
  \citenamefont {Page}, \citenamefont {Pryke}, \citenamefont {Reichardt},
  \citenamefont {Ruhl}, \citenamefont {Sehgal}, \citenamefont {Seljak},
  \citenamefont {Sievers}, \citenamefont {Silverstein}, \citenamefont {Slosar},
  \citenamefont {Smith}, \citenamefont {Spergel}, \citenamefont {Staggs},
  \citenamefont {Stark}, \citenamefont {Stompor}, \citenamefont {Vieregg},
  \citenamefont {Wang}, \citenamefont {Watson}, \citenamefont {Wollack},
  \citenamefont {Wu}, \citenamefont {Yoon},\ and\ \citenamefont
  {Zahn}}]{abazajian_neutrino_2013}%
  \BibitemOpen
  \bibfield  {author} {\bibinfo {author} {\bibfnamefont {K.~N.}\ \bibnamefont
  {Abazajian}}, \bibinfo {author} {\bibfnamefont {K.}~\bibnamefont {Arnold}},
  \bibinfo {author} {\bibfnamefont {J.}~\bibnamefont {Austermann}}, \bibinfo
  {author} {\bibfnamefont {B.~A.}\ \bibnamefont {Benson}}, \bibinfo {author}
  {\bibfnamefont {C.}~\bibnamefont {Bischoff}}, \bibinfo {author}
  {\bibfnamefont {J.}~\bibnamefont {Bock}}, \bibinfo {author} {\bibfnamefont
  {J.~R.}\ \bibnamefont {Bond}}, \bibinfo {author} {\bibfnamefont
  {J.}~\bibnamefont {Borrill}}, \bibinfo {author} {\bibfnamefont
  {E.}~\bibnamefont {Calabrese}}, \bibinfo {author} {\bibfnamefont {J.~E.}\
  \bibnamefont {Carlstrom}}, \bibinfo {author} {\bibfnamefont {C.~S.}\
  \bibnamefont {Carvalho}}, \bibinfo {author} {\bibfnamefont {C.~L.}\
  \bibnamefont {Chang}}, \bibinfo {author} {\bibfnamefont {H.~C.}\ \bibnamefont
  {Chiang}}, \bibinfo {author} {\bibfnamefont {S.}~\bibnamefont {Church}},
  \bibinfo {author} {\bibfnamefont {A.}~\bibnamefont {Cooray}}, \bibinfo
  {author} {\bibfnamefont {T.~M.}\ \bibnamefont {Crawford}}, \bibinfo {author}
  {\bibfnamefont {K.~S.}\ \bibnamefont {Dawson}}, \bibinfo {author}
  {\bibfnamefont {S.}~\bibnamefont {Das}}, \bibinfo {author} {\bibfnamefont
  {M.~J.}\ \bibnamefont {Devlin}}, \bibinfo {author} {\bibfnamefont
  {M.}~\bibnamefont {Dobbs}}, \bibinfo {author} {\bibfnamefont
  {S.}~\bibnamefont {Dodelson}}, \bibinfo {author} {\bibfnamefont
  {O.}~\bibnamefont {Dore}}, \bibinfo {author} {\bibfnamefont {J.}~\bibnamefont
  {Dunkley}}, \bibinfo {author} {\bibfnamefont {J.}~\bibnamefont {Errard}},
  \bibinfo {author} {\bibfnamefont {A.}~\bibnamefont {Fraisse}}, \bibinfo
  {author} {\bibfnamefont {J.}~\bibnamefont {Gallicchio}}, \bibinfo {author}
  {\bibfnamefont {N.~W.}\ \bibnamefont {Halverson}}, \bibinfo {author}
  {\bibfnamefont {S.}~\bibnamefont {Hanany}}, \bibinfo {author} {\bibfnamefont
  {S.~R.}\ \bibnamefont {Hildebrandt}}, \bibinfo {author} {\bibfnamefont
  {A.}~\bibnamefont {Hincks}}, \bibinfo {author} {\bibfnamefont
  {R.}~\bibnamefont {Hlozek}}, \bibinfo {author} {\bibfnamefont
  {G.}~\bibnamefont {Holder}}, \bibinfo {author} {\bibfnamefont {W.~L.}\
  \bibnamefont {Holzapfel}}, \bibinfo {author} {\bibfnamefont {K.}~\bibnamefont
  {Honscheid}}, \bibinfo {author} {\bibfnamefont {W.}~\bibnamefont {Hu}},
  \bibinfo {author} {\bibfnamefont {J.}~\bibnamefont {Hubmayr}}, \bibinfo
  {author} {\bibfnamefont {K.}~\bibnamefont {Irwin}}, \bibinfo {author}
  {\bibfnamefont {W.~C.}\ \bibnamefont {Jones}}, \bibinfo {author}
  {\bibfnamefont {M.}~\bibnamefont {Kamionkowski}}, \bibinfo {author}
  {\bibfnamefont {B.}~\bibnamefont {Keating}}, \bibinfo {author} {\bibfnamefont
  {R.}~\bibnamefont {Keisler}}, \bibinfo {author} {\bibfnamefont
  {L.}~\bibnamefont {Knox}}, \bibinfo {author} {\bibfnamefont {E.}~\bibnamefont
  {Komatsu}}, \bibinfo {author} {\bibfnamefont {J.}~\bibnamefont {Kovac}},
  \bibinfo {author} {\bibfnamefont {C.-L.}\ \bibnamefont {Kuo}}, \bibinfo
  {author} {\bibfnamefont {C.}~\bibnamefont {Lawrence}}, \bibinfo {author}
  {\bibfnamefont {A.~T.}\ \bibnamefont {Lee}}, \bibinfo {author} {\bibfnamefont
  {E.}~\bibnamefont {Leitch}}, \bibinfo {author} {\bibfnamefont
  {E.}~\bibnamefont {Linder}}, \bibinfo {author} {\bibfnamefont
  {P.}~\bibnamefont {Lubin}}, \bibinfo {author} {\bibfnamefont
  {J.}~\bibnamefont {McMahon}}, \bibinfo {author} {\bibfnamefont
  {A.}~\bibnamefont {Miller}}, \bibinfo {author} {\bibfnamefont
  {L.}~\bibnamefont {Newburgh}}, \bibinfo {author} {\bibfnamefont {M.~D.}\
  \bibnamefont {Niemack}}, \bibinfo {author} {\bibfnamefont {H.}~\bibnamefont
  {Nguyen}}, \bibinfo {author} {\bibfnamefont {H.~T.}\ \bibnamefont {Nguyen}},
  \bibinfo {author} {\bibfnamefont {L.}~\bibnamefont {Page}}, \bibinfo {author}
  {\bibfnamefont {C.}~\bibnamefont {Pryke}}, \bibinfo {author} {\bibfnamefont
  {C.~L.}\ \bibnamefont {Reichardt}}, \bibinfo {author} {\bibfnamefont {J.~E.}\
  \bibnamefont {Ruhl}}, \bibinfo {author} {\bibfnamefont {N.}~\bibnamefont
  {Sehgal}}, \bibinfo {author} {\bibfnamefont {U.}~\bibnamefont {Seljak}},
  \bibinfo {author} {\bibfnamefont {J.}~\bibnamefont {Sievers}}, \bibinfo
  {author} {\bibfnamefont {E.}~\bibnamefont {Silverstein}}, \bibinfo {author}
  {\bibfnamefont {A.}~\bibnamefont {Slosar}}, \bibinfo {author} {\bibfnamefont
  {K.~M.}\ \bibnamefont {Smith}}, \bibinfo {author} {\bibfnamefont
  {D.}~\bibnamefont {Spergel}}, \bibinfo {author} {\bibfnamefont {S.~T.}\
  \bibnamefont {Staggs}}, \bibinfo {author} {\bibfnamefont {A.}~\bibnamefont
  {Stark}}, \bibinfo {author} {\bibfnamefont {R.}~\bibnamefont {Stompor}},
  \bibinfo {author} {\bibfnamefont {A.~G.}\ \bibnamefont {Vieregg}}, \bibinfo
  {author} {\bibfnamefont {G.}~\bibnamefont {Wang}}, \bibinfo {author}
  {\bibfnamefont {S.}~\bibnamefont {Watson}}, \bibinfo {author} {\bibfnamefont
  {E.~J.}\ \bibnamefont {Wollack}}, \bibinfo {author} {\bibfnamefont
  {W.~L.~K.}\ \bibnamefont {Wu}}, \bibinfo {author} {\bibfnamefont {K.~W.}\
  \bibnamefont {Yoon}}, \ and\ \bibinfo {author} {\bibfnamefont
  {O.}~\bibnamefont {Zahn}},\ }\href {http://arxiv.org/abs/1309.5383}
  {\bibfield  {journal} {\bibinfo  {journal} {arXiv:1309.5383 [astro-ph,
  physics:hep-ph]}\ } (\bibinfo {year} {2013})},\ \bibinfo {note} {arXiv:
  1309.5383}\BibitemShut {NoStop}%
\bibitem [{\citenamefont {{Ng}}\ and\ \citenamefont
  {{Beacom}}(2014)}]{ng_2015_cascades}%
  \BibitemOpen
  \bibfield  {author} {\bibinfo {author} {\bibfnamefont {K.~C.~Y.}\
  \bibnamefont {{Ng}}}\ and\ \bibinfo {author} {\bibfnamefont {J.~F.}\
  \bibnamefont {{Beacom}}},\ }\href {\doibase 10.1103/PhysRevD.90.065035}
  {\bibfield  {journal} {\bibinfo  {journal} {\prd}\ }\textbf {\bibinfo
  {volume} {90}},\ \bibinfo {eid} {065035} (\bibinfo {year} {2014})},\ \Eprint
  {http://arxiv.org/abs/1404.2288} {arXiv:1404.2288 [astro-ph.HE]} \BibitemShut
  {NoStop}%
\end{thebibliography}%

\end{document}